\documentstyle[11pt,newpasp,twoside,psfig]{article}
\index{Chandra}
\index{optical}
\index{pulsars!J1124--5916}
\index{pulsar wind nebulae!G292.0+1.8}
\index{pulsar wind nebulae!interactions with supernova remnants}

\def\gtwo{G292.0+1.8}
\def\psr{PSR~J1124$-$5916}
\def\astroe{{\em Astro-E2}}
\def\conx{{\em Constellation-X Facility}}
\def\chandra{{\em Chandra}}
\def\cxo{{\em Chandra X-Ray Observatory}}

\def\edcomment#1{\iffalse\marginpar{\raggedright\sl#1\/}\else\relax\fi}
\reversemarginpar

\begin{document}
\title{An X-Ray Pulsar, Metal-rich Ejecta, and Shocked 
Ambient Medium in the Supernova Remnant \gtwo}
\author{John P. Hughes, Robert B. Friedman}
\affil{Rutgers University}
\author{Patrick Slane}
\affil{Harvard-Smithsonian Center for Astrophysics}
\author{Sangwook Park}
\affil{Pennsylvania State University}

\begin{abstract}

We report the discovery of pulsed X-ray emission from the compact
object CXOU J112439.1$-$591620 within the Galactic supernova remnant \gtwo\
using the High Resolution Camera on the \cxo.  The X-ray period is
consistent with the extrapolation of the radio period and spindown
rate of \psr. The X-ray pulse is single peaked and broad. There is no
optical counterpart to a limit of $M_V \sim 26$. The pressure in the
pulsar wind nebula is considerably less than that in the
reverse-shock--heated ejecta and circumstellar medium, indicating that
the reverse shock has not yet begun to interact with the nebula.

\end{abstract}

\section{Introduction}

\gtwo\ is a young, oxygen-rich supernova remnant (SNR) in the southern
sky. The recent discovery of a pulsar wind nebula (PWN) (Hughes et
al. 2001) and 135-ms period radio pulsar (Camilo et al.~2002) have
greatly increased interest in this object, which, because of its
youth, ejecta-dominated nature, and relative closeness, is sometimes
considered the southern analog of Cassiopeia A.  
Here we
report on the nature of the pulsar and discuss the
evolutionary state of the SNR and PWN, especially as regards their
interaction.

%

\section{X-Ray Pulsar}

The compact X-ray source CXOU J112439.1$-$591620 in \gtwo\ was
observed with the \chandra\ high resolution camera (HRC) in timing
mode for 50 ks on 2001 July 14.  No pulse was detected in a coherent
FFT of the entire light curve for the events (1324 in total) extracted
from within a $2^{\prime\prime}$ radius of the point source. A search
at pulse frequencies near the radio parameters yielded a significant
detection of an X-ray pulsar (Figure 1), confirming that the radio
pulsar, X-ray pulsar, and compact central X-ray source are one and
the same.  Agreement between the ages of the pulsar and SNR, make a
virtually ironclad case for an association between \psr\ and
SNR \gtwo.

The X-ray pulse is single peaked and broad (Figure 2). Of the total
number of events in the light curve, only about 130 are pulsed.  The
X-ray image in the vicinity of the pulsar is complex: in addition to
an unresolved source there is a small, extended, elliptically-shaped
nebula.  In fact the nebula dominates the flux from this region; only
approximately 160 events (time-averaged) come from the unresolved
pulsar. Comparing the pulsed to the time-averaged flux indicates that
the X-ray pulsar is highly pulsed, $>$65\%.  A more complete report on
the properties of the X-ray pulsar and its compact nebula can be found
in Hughes et al.~(2003).

\begin{figure}
\begin{center}
 \psfig{file=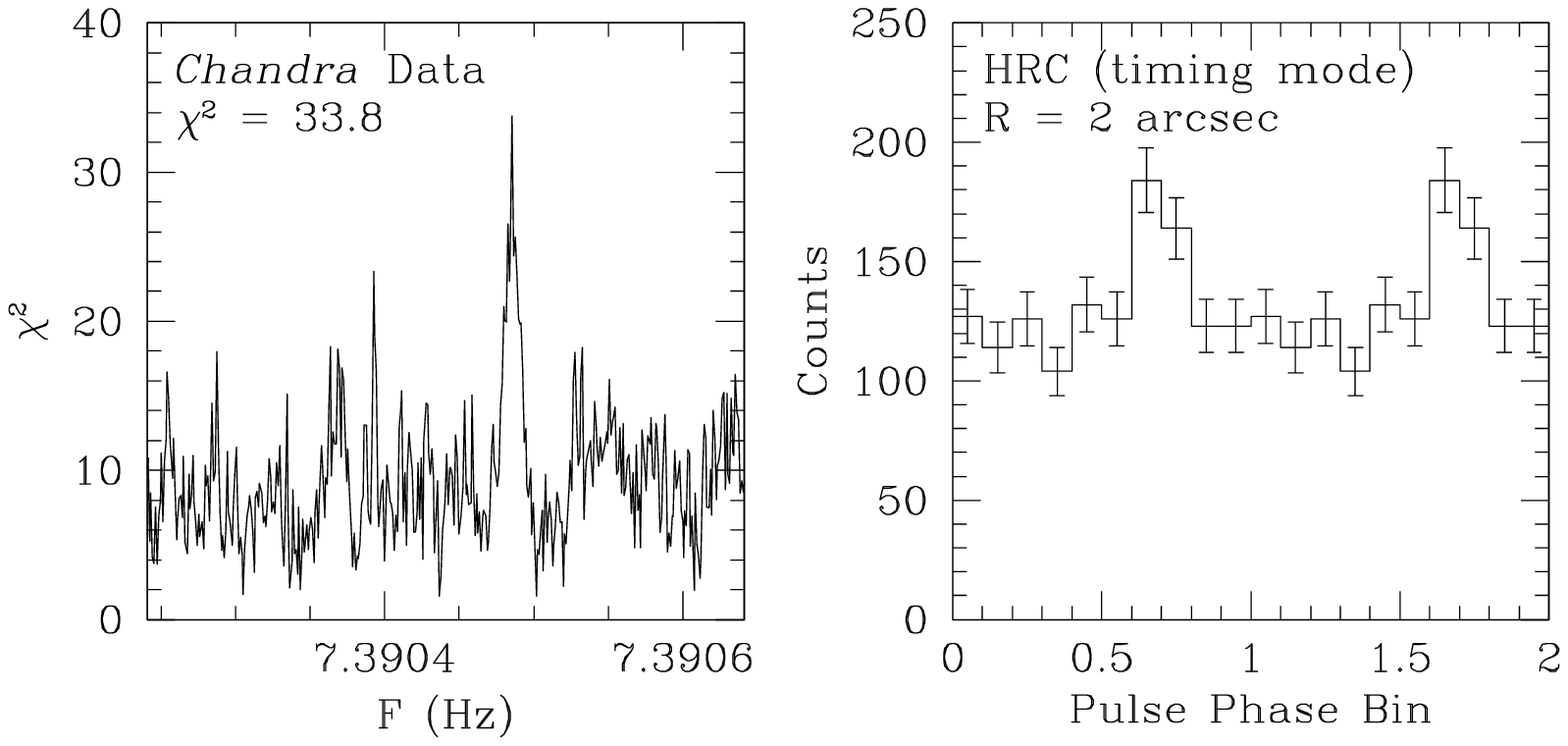,width=\textwidth}
\end{center}
\vskip -0.25truein
\caption{{\em(Left)} $\chi^2$ vs.\ trial frequency for the compact central X-ray source
in \gtwo. The peak frequency is consistent with that of  \psr.}
\caption{{\em(Right)} Pulse phase X-ray light curve.}
\end{figure}

\section{Optical Limits on \psr}

In a search for an optical counterpart, the pulsar was observed at the
CTIO Blanco 4-m telescope with the MOSAIC II camera on 2002 April
17. Exposures were taken in one narrow-band H$\alpha$ and four
broadband filters for approximately 1 hour each.  Standard reductions
were carried out and the images were registered to the celestial
sphere.  No optical counterpart was detected (Figure 3) down to limits
in B, V, R, and I of 26.8, 26.4, 26.2, and 25.5, respectively. The
intrinsic optical luminosity for PSR 1124$-$5916, accounting for
distance and extinction, is much below the luminosities of the three
young pulsars Crab, PSR B0540$-$69 and PSR B1509$-$58, although the
limit is still above the optical luminosity of PSR B0833$-$45 in
Vela. In terms of the efficiency for converting spin down energy into
optical luminosity, these new limits for PSR 1124$-$5916 imply a
rather low value: $\epsilon = L_V / \dot E < 1.7\times 10^{-7}$.
Again this falls far below the values found for the three young
isolated pulsars in SNRs, but is an order of magnitude or so higher
than the value of $\epsilon$ for the Vela pulsar (Caraveo, Mereghetti,
\& Bignami 1994).

\begin{figure}
\begin{center}
 \hbox{
 \psfig{file=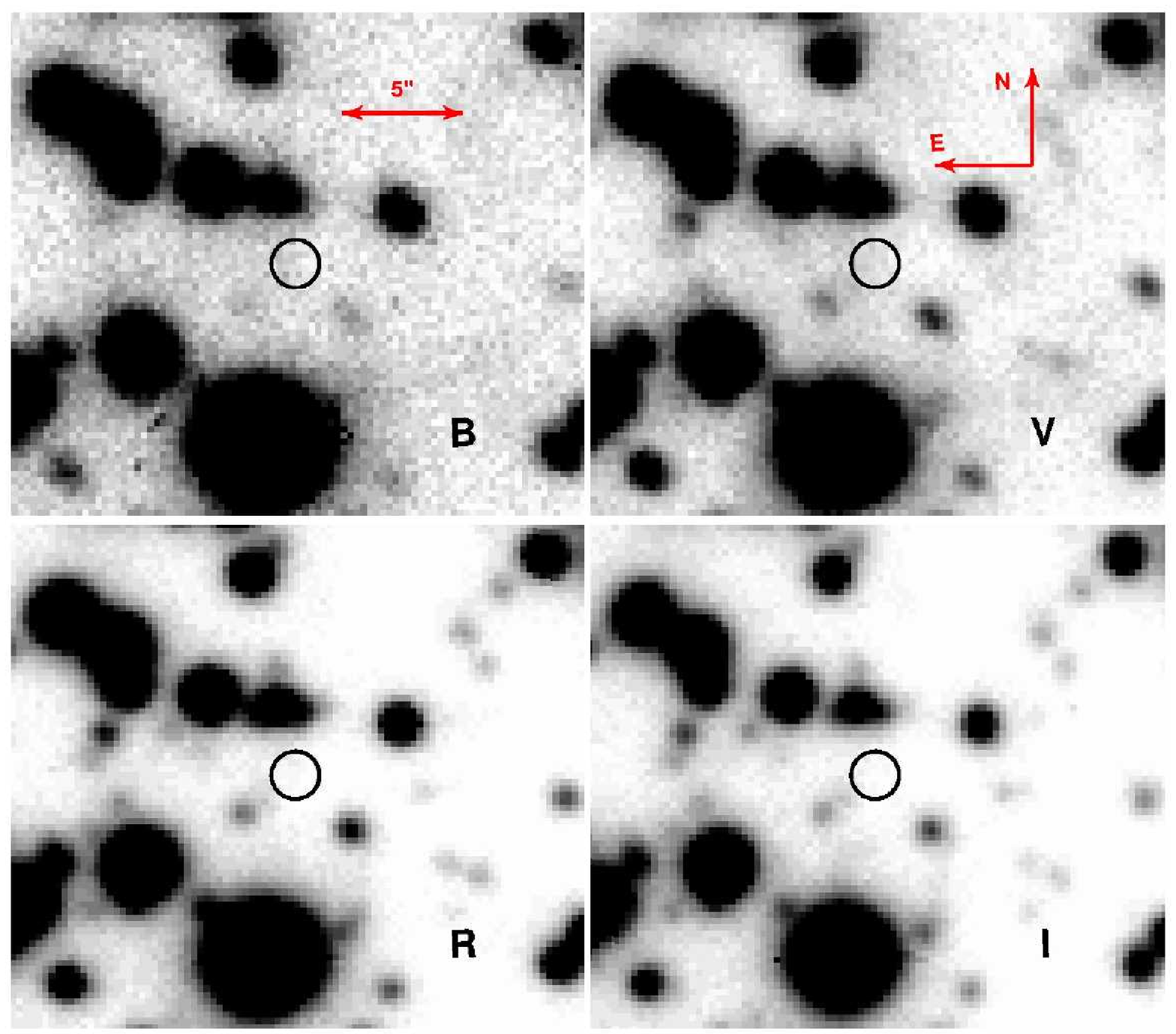,width=2.7truein}
 \psfig{file=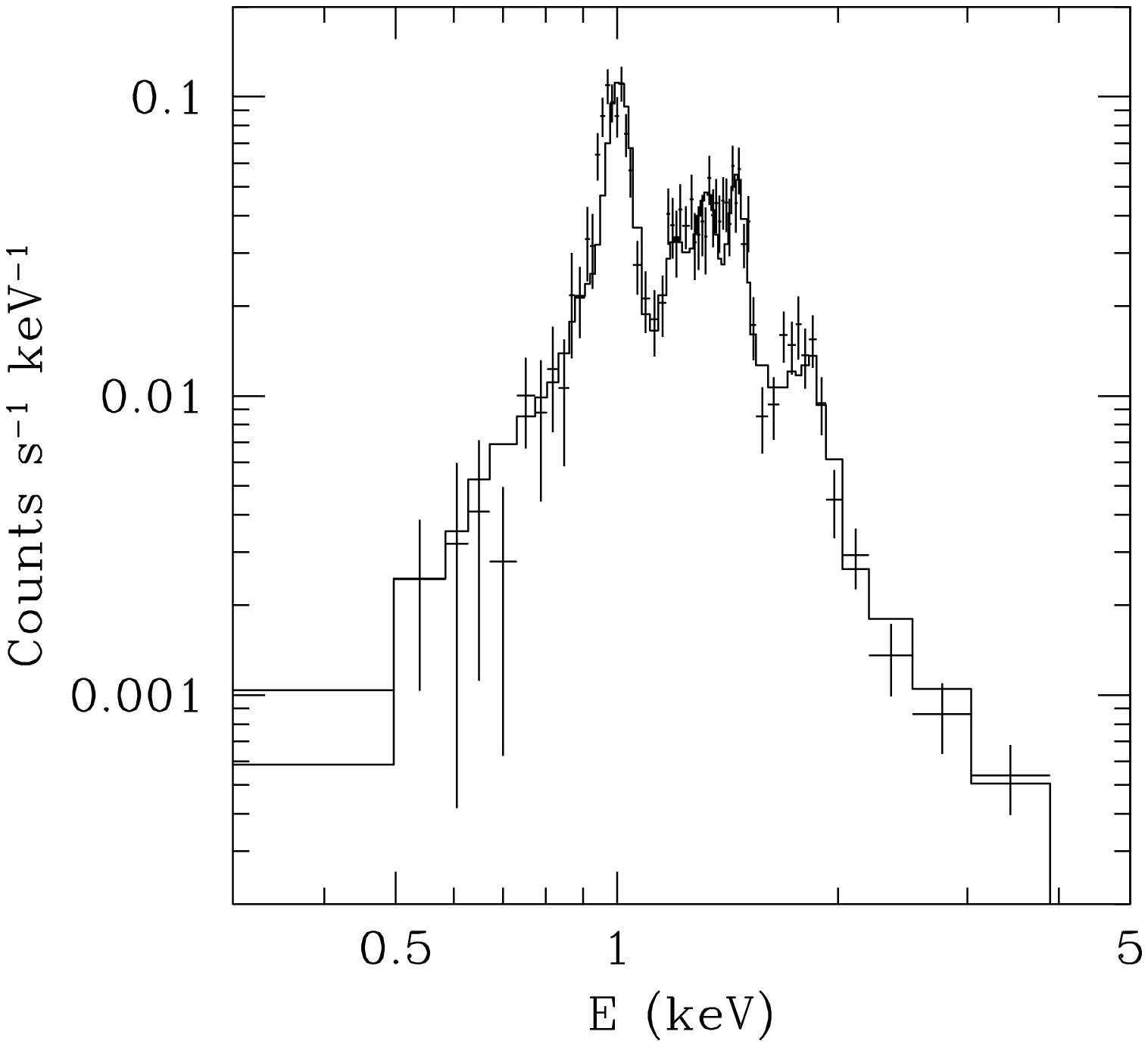,width=2.5truein}
 }
\end{center}
\vskip -0.25truein
\caption{
{\em(Left)} Broadband optical images in the vicinity of \psr, whose
position is denoted by the circle. North is up; east is to left.
}
\caption{
{\em(Right)} \chandra\ ACIS-S spectrum of an ejecta knot
}

\end{figure}

\section{Dynamical State of the PWN and SNR}

One of the most important issues regarding the evolutionary state of
\gtwo\ concerns the relationship between the PWN, the free-streaming
ejecta, and the reverse shock. By examining the pressures in the
various regions of the remnant we have obtained some insight into this
question.

The pressure in the pulsar wind nebula has been estimated in two
rather different ways (Hughes et al.~2003): using the location of the
pulsar termination shock and from the minimum energy condition applied
to the radio synchrotron emission. Values obtained were
$1.3\times 10^{-10}$ dyne cm$^{-2}$ (minimum energy) and $2.6\times
10^{-9}$ dyne cm$^{-2}$ (termination shock).  Although it is possible
that the pressure is as high as the latter value, in practice this
would require the nebula to be almost entirely particle dominated with
an energy content 40 times that of the minimum energy condition. Any
value of pressure much higher than this would be very difficult to
accommodate given what we currently know about the age and energetics
of \gtwo. It therefore seems reasonable to us to conclude that the
true pressure in the PWN lies at or below $\sim$$10^{-9}$ dyne cm$^{-2}$.

Recently, a study of the thermal properties of \gtwo\ has been carried
out by Park et al.~(2003). Regions of normal composition,
corresponding to shocked interstellar or circumstellar medium
(ISM/CSM), were identified based on near-solar fitted abundance
ratios.  Using the densities and temperatures derived from the
spectral fits for these regions, an emission geometry, and the
assumption of equal electron and ion temperatures, an estimate for the
thermal pressure of $8.8\times 10^{-8}$ dyne cm$^{-2}$ was obtained.
There are numerous knots of X-ray emitting ejecta throughout the
remnant.  Figure 4 shows the \chandra\ ACIS-S spectrum of one such
knot located slightly south of the pulsar.
%
%
The knot shows strong emission lines of Ne ($\sim$1 keV), Mg
($\sim$1.3 keV), and Si ($\sim$1.8 keV).  The spectrum can be fully
described with emission contributions from only these three species
and oxygen; the data do not require the presence of any other
elemental species.  We identify this as a knot of pure ejecta from the
hydrostatic burning layers of the $\sim$$25\,M_\odot$ progenitor star
(Park et al.~2003).  In a similar manner to before we estimate the
thermal pressure in this knot to be $5.9\times 10^{-9}$ dyne
cm$^{-2}$.  Although there are considerable uncertainties in these
pressure estimates, the most critical assumptions we have made, e.g.,
that the plasma is unclumped and that electrons and ions share the
same temperature, would tend to result in higher pressure estimates if
relaxed.

The pressure in the shocked ejecta and ISM/CSM is at least a factor of
two and more likely one to two orders of magnitude higher than the
pressure in the PWN.  This is strong evidence that the reverse shock
has not yet begun to encounter the PWN. Furthermore, our ISM/CSM,
ejecta, and PWN pressures are in remarkably good numerical agreement
with dynamical models for a PWN interacting with freely expanding
ejecta (see figure 4 of van der Swaluw et al.~2001).  Finding evidence
for the interaction between the PWN and the free streaming ejecta
would be an important confirmation of this picture.

There should also be a reservoir of cold unshocked ejecta between the
PWN and the reverse shock.  One way to detect and investigate the
composition of this material would be through absorption line studies
of, for example, cold Fe or Si against the hard continuum emission of
the PWN.  This may be possible to do using \astroe\ and should become
routine with future higher sensitivity missions like the \conx.


\acknowledgments The authors thank F.\ Camilo, B.\ Gaensler, M.\ Juda, 
P.\ Ghavamian,  D.\ Burrows, J.\ Nousek, G.\ Garmire,
and F.\ Winkler for help with various aspects of this research or
for contributions to the presentation. Financial support was partially
provided by \chandra\ grant GO1-2052X.

\end{document}